\documentclass[
reprint,
nofootinbib,
 amsmath,amssymb,
 aps,
]{revtex4-2}

\usepackage{graphicx}
\usepackage{dcolumn}
\usepackage{xcolor} 
\usepackage{amsmath}
\usepackage{braket}
\usepackage{float}
\usepackage{booktabs}
\usepackage{bm}
\usepackage[normalem]{ulem}

\definecolor{mediumgreen}{RGB}{0,170,0}

\definecolor{mediumblue}{RGB}{0,0,170}

\begin{document}

\preprint{APS/123-QED}

\title{Multiphoton Electronic Coherences in Molecules}

\author{Loc Thi-Hoang Ngo$^1$}
\author{Brian Kaufman$^2$}
\author{Varun Makhija$^3$}
\author{Ryan Cloughly$^3$}
\author{Javin Ohara$^4$}
\author{Julia Codere$^1$}
\author{Scott Liguori$^1$}
\author{Sameer Singh$^1$}
\author{Martin Cohen$^1$}
\author{Matt Bain$^2$}
\author{Brett Pearson$^5$}
\author{Ruaridh Forbes$^6$}
\author{Thomas Weinacht$^1$}
\email{thomas.weinacht@stonybrook.edu}

\affiliation{$^{1}$Department of Physics and Astronomy, Stony Brook University, Stony Brook NY 11794-3800, USA.}
\affiliation{$^{2}$Linac Coherent Light Source, SLAC National Accelerator Laboratory, Menlo Park, CA 94025, USA.}
\affiliation{$^{3}$Department of Chemistry and Physics, University of Mary Washington, 1301 College Ave, Fredericksburg, VA 22401, USA.}
\affiliation{$^{4}$Department of Physics, Stanford University, Stanford CA, 94305} 
\affiliation{$^{5}$Department of Physics and Astronomy, Dickinson College, Carlisle, PA 17013, USA.}
\affiliation{$^{6}$Department of Chemistry, University of California, Davis, One Shields Ave, Davis, CA 95616, USA.}
\begin{abstract}
We investigate multiphoton electronic coherences in a series of gas--phase atomic and molecular systems with increasing complexity. Our measurements are driven by pairs of phase-locked, deep-ultraviolet pulses with independent control of the relative phase and time delay. By combining the ionization yields measured at specific relative phases  between the pulses, we can isolate different multiphoton orders and enable measurements of coherences between specific pairs of electronic states. We measure electronic coherences between states separated by zero-photon (i.e. within our laser bandwidth), one-photon and two-photons, resulting in signals with modulation periods ranging from 430~attoseconds to several femtoseconds. Fast electronic dynamics are tracked simultaneously with the accompanying nuclear motion. Additionally, we derive analytic expressions for the measured ionization yield as a function of pump–probe delay and demonstrate their use as a quantitative tool for isolating individual photon-order contributions within the total yield.
\end{abstract}

\maketitle


\section{\label{sec:level1}Introduction}
Two central objectives of attosecond science are to observe and control the motion of electrons in matter on their natural timescales \cite{corkum2007attosecond, krausz2009attosecond, palacios2020quantum,okino2015direct, goulielmakis2010real}. Electronic dynamics initiated by photoexcitation underlie a broad range of fundamental processes, including charge migration, energy redistribution, ionization, and the early stages of chemical transformation \cite{worner2017charge, folorunso2021molecular, calegari2014ultrafast, nisoli2017attosecond, kraus2015measurement, matselyukh2022decoherence,vester2023role, ramasesha2016real, despre2018charge}. Given that the electron density distribution determines the forces on the nuclei, and that electrons respond to external perturbations on much shorter timescales than nuclei, electron dynamics can determine the subsequent nuclear motion involved in molecular transformation. Understanding how electronic wave functions evolve following photoexcitation is therefore essential for uncovering the fundamental mechanisms governing molecular photophysics and photochemistry, particularly during the initial stages of photoinduced processes before substantial nuclear motion occurs \cite{ruberti2022quantum}.

Despite remarkable advances in ultrafast spectroscopy, the investigation of molecular electronic dynamics remains experimentally challenging. Attosecond electronic dynamics are difficult to drive because one needs to create coherent superpositions of electronic eigenstates. Achieving this requires ultrashort laser pulses with broad spectral bandwidths. While attosecond XUV pulses provide exceptional temporal resolution, their photon energies are generally too high to directly access neutral excited states of molecules. Consequently, the direct preparation of a coherent superposition of multiple neutral eigenstates requires an extremely broadband UV excitation source. These wave packets are also difficult to measure because the generation of attosecond light sources capable of delivering sufficient intensity to produce measurable ionization yields remains technically demanding \cite{guo2024experimental, kretschmar2024compact}.  Furthermore, electronic wave packets  undergo rapid dephasing, state mixing, and coupling to nuclear motion \cite{hwang2004electronic, kamisaka2006ultrafast, vacher2015electron,Vacher2017ElectronDynamics, kaufman2023long, kaufman2024long, mougol2024direct}, further complicating the interpretation of experimental observables. 

In this work, we make use of phase-locked pairs of short UV pulses and low-order multiphoton excitation to get around those difficulties. This enables excitation to multiple electronic and vibrational states at each photon order, along with the simultaneous excitation to multiple photon orders. Recent work considered simple electronic wave packets formed by single photon excitation from the ground state which prepared a coherent superposition of ground and excited electronic states \cite{ngo2026molecular}. Here, we consider excitation through both one- and two-photon absorption, involving multiple electronic states at each photon order. Measurements in atomic xenon provide a rigorous test of the experimental apparatus and protocol, given the well-characterized atomic energy levels and states. Having established the reliability of the approach, we extend its application to molecular systems, demonstrating the ability to follow long-lived electronic coherences in the face of decoherence driven by coupling with nuclear degrees of freedom.

\section{Analytic Model for Multiplexed Multiphoton Spectroscopy} \label{sec:alphas}
We model the experimental signal by applying time-dependent perturbation theory to a wave packet, or a coherent superposition of molecular eigenstates $\ket{\psi_j}$;
\begin{equation}
    \ket{\Psi(t)}
    =\sum_{j}a_{j}e^{-iE_jt/\hbar}\ket{\psi_j}
    \label{eq:wkpt-es}
\end{equation}
where $\ket{\psi_j}$ is the $j^{\text{th}}$ eigenstate of the molecule (including both electronic and nuclear degrees of freedom), $a_j$ and $E_j$ are the amplitude and energy of the $j^{\text{th}}$ state, respectively. We assume that a pump pulse of the form $E_{\mathrm{pump}}(t)=E_0(t)(e^{i\omega_\ell t}+c.c.)$ creates such a wave packet, where $E_0(t)=E_0e^{-t^2/\sigma^2}$, with $E_0$ representing the amplitude of the pulse, $\sigma$ the width, and $\omega_\ell$ the laser frequency.  This wave packet evolves for a time $\tau$, at which point it interacts with a probe pulse of the form $E_{\mathrm{probe}}(t)=E_0(t-\tau)(e^{i(\omega_\ell (t-\tau)+\phi)}+c.c)$, where $\phi$ is the phase of the probe pulse relative to the pump (both pulses have the same envelope and central frequency).
In the limit where the probe pulse is shorter than the timescale for nuclear dynamics, the Born-Oppenheimer approximation can be assumed to hold during the probe pulse, in which case the eigenstates $\ket{\psi_j}$ are the molecular electronic states. The ionization yield that follows from Eq.~\ref{eq:wkpt-es} can therefore also be analyzed in terms of electronic states, provided nuclear motion is negligible during the probe pulse. Earlier work made use of a Born-Oppenheimer expansion to relate the measured ionization yield to the underlying wave packet dynamics \cite{ngo2026molecular,kaufman2023long,kaufman2024long}.

The measurement observable consists of the ionization yield as a function of delay and relative phase between the pump and probe pulses. The ionization yield can be directly related to the quantum mechanical wave function and the density matrix elements.  A detailed derivation is presented in the Appendix~\ref{sec:app}, which provides the following expression for the ionization yield:
\begin{equation}
\begin{split}
&Y(\tau,\phi)=\\
&\sum_{p=-\infty}^{\infty}e^{ip\phi}\sum_{q\geq p} \sum_{\gamma\gamma'}...\sum_{jj'}W^{(q)}_{cj}W^{(q-p)*}_{cj'}Q^{(q)}_{cj}Q^{(q-p)*}_{cj'}\rho_{jj'}(\tau)
\label{eq:ptsignalmain}
\end{split}
\end{equation}
where we have used the abbreviations,
\begin{align}
    W^{(q)}_{cj} &\equiv w^{(q)}_{cj}(\Delta E_{c\gamma},...,\Delta E_{\beta k},\Delta E_{k j}) \\
    Q^{(q)}_{cj} &\equiv \bra{\psi_c}\hat{\mu}\ket{\psi_\gamma}...\bra{\psi_\beta}\hat{\mu}\ket{\psi_k}\bra{\psi_k}\hat{\mu}\ket{\psi_j}.
    \label{eq:Q_defmain}
\end{align}
Here the $\rho_{jj'}(\tau)$ are the pump-probe delay ($\tau$) dependent density matrix elements. The sum over $q$ represents the sum over ionization pathways with order $q$. The sum over $p$ represents the sum over the complementary $q-p$ order ionization pathways for pairs of states separated in energy by $p\hbar \omega_l$ photons.  The sum over $k$ represents the sum over all of the one-photon intermediate states in the $q$ photon ionization pathway, and the sum over $k'$ represents the sum over all of the one-photon intermediate states in the complementary $q-p$ photon ionization pathway. The sum over $j$ and $j'$ is the sum over all states $\ket{\psi_j}$ and $\ket{\psi_{j'}}$ in the wave packet created by the pump pulse (see Eq.~\ref{eq:wkpt-es}).  The $Q^{(q)}_{cj}$ are a chain of $q$-photon photoionization dipole matrix elements initiating in the state $\ket{\psi_j}$, facilitating subsequent transitions through the intermediate states $\ket{\psi_\gamma}...\ket{\psi_\beta},\ket{\psi_k}$, and terminating in the cation state $\ket{\psi_c}$. We note that there is a nested sum at each photon order, as shown in Eq. \ref{eq:Q_defmain}, which can require separate sums such as the one over $k$ and $k'$ in Eq. \ref{eq:ptsignalmain}. The $W_{cj}^{(q)}$ are the corresponding `window' functions that enforce the conservation of energy for each transition. An explicit expression and further discussion of these functions are provided in the Appendix~\ref{sec:app}.  

The ionization yield in Eq.~\ref{eq:ptsignalmain} can be viewed as an expansion in all pairs of quantum pathways to the final state $\ket{\psi_c}$, each pathway initiating from a pair of states $\ket{\psi_j}$ and $\ket{\psi_{j'}}$ that reach $\ket{\psi_c}$ by the absorption of $q$ and $q'=q-p$ photons, respectively. In this sense there is some similarity to the RABBITT technique, which uses interfering pathways to reconstruct the relative phase between harmonics in an attosecond pulse train \cite{paul2001observation}. The phase of each term in the expansion depends on the difference in photon order between the pathways which, we define as $p=q-q'$. We can then rewrite Eq.~\ref{eq:ptsignalmain} as,
\begin{equation}
Y(\tau,\phi)=\sum_{p=-\infty}^{\infty}\alpha_p(\tau)e^{ip\phi},
\label{eq:alpha-signal}
\end{equation}
where $\alpha_p(\tau)$ is defined as
\begin{equation}
\alpha_p(\tau) = \sum_{q\geq p}\sum_{\gamma\gamma'}...\sum_{jj'}W^{(q)}_{cj}W^{(q-p)*}_{cj'}Q^{(q)}_{cj}Q^{(q-p)*}_{cj'}\rho_{jj'}(\tau).
\label{eq:alpha-rho}
\end{equation}
Equation~\ref{eq:alpha-signal} is evidently a complex Fourier series in the photon order difference $p$ with conjugate variable $\phi$. Therefore, with a sufficient set of measurements $Y(\tau,\phi)$, a set of individual photon order signals $\alpha_p(\tau)$ can be easily obtained using the orthogonality and completeness of the functions $e^{ip\phi}$ in the range $\phi = \{-\pi,\pi\}$:
\begin{equation}
\alpha_p(\tau)=\frac{1}{2\pi}\int_{-\pi}^{\pi}e^{-ip\phi}Y(\tau,\phi)d\phi.
\label{eq:get_alpha}
\end{equation}

In practice, the inverse Discrete Fourier Transform (DFT) with respect to $\phi$ of the measured $Y(\tau,\phi)$ at each $\tau$ will provide the set of $\alpha_p(\tau)$. The minimum number of different $\phi$ measurements of the signal required for a successful inversion is determined by the Nyquist condition for a given maximum photon order difference $p_{\text{max}}$. From Eq.~\ref{eq:alpha-signal} we see that the $p$ are the discrete `frequencies' of the Fourier series. Therefore, to properly sample a frequency $p_{\text{max}}$ the Nyquist condition requires a sampling frequency greater than  $2p_{\text{max}}$, and so a minimum sampling frequency of $2p_{\text{max}} + 1$. For instance, if we expect pathways that can reach the same final state and differ by at most 3 photons, then $p_{\text{max}} = 7$ and at least seven measurements of $Y(\tau,\phi)$ are needed for equally spaced values $\phi$ between $-\pi$ and $\pi$.

Equivalently, Eq.~\ref{eq:alpha-signal} can be viewed as a set of linear equations in the $2p_{\text{max}}+1$ unknowns $\alpha_p(\tau)$. Formulated as a matrix equation this becomes $\vec{Y}(\tau)=\hat{M}\vec{\alpha}(\tau)$, where $\vec{Y}(\tau)$ is a column vector populated with measurements of $Y(\tau,\phi)$ for different values of $\phi$ at each delay, $\hat{M}$ is a matrix of the values $e^{ip\phi}$ with $2p_{\text{max}}+1$ rows and one column for each $\phi$ measurement, and $\vec{\alpha}(\tau)$ is a column vector with $2p_{\text{max}}+1$ columns. Again, if the number of phase measurements is $2p_{\text{max}}+1$, and the phases are measured in the range $\{-n\pi,n\pi\}$ for any integer $n$, $\hat{M}$ will be a uniquely invertible square matrix with orthogonal columns providing a unique solution for $\vec{\alpha}(\tau)$ at each delay. 

Each photon-order-resolved signal $\alpha_p(\tau)$ can be related to a $p^{\text{th}}$-order time-domain spectroscopy. The \emph{zero-photon signal}, $\alpha_0(\tau)$,  results only from pairs of wave packet states that have energy gaps spanned by the bandwidth of the probe pulse - i.e. by ``zero photons". This is typically what is measured in the electron-kinetic-energy-integrated, time-resolved photoelectron spectroscopy signal (TRPES), assuming one-photon ionization of the excited wave packet. As has been previously demonstrated, this will thus contain information on the population of individual electronic states, and on coherences between different electronic states that ionize to the same final state with one photon~\cite{stolow2008,gregory2022,morrigan2023}.

The \emph{one-photon signal}, $\alpha_1(\tau)$, results from pairs of wave packet states with energy gaps matching the energy of a probe photon that can be ionized to the same state in the continuum. From Eq.~\ref{eq:alpha-rho} we see that $\alpha_1(\tau)$ is composed specifically of coherences $\rho_{jj'}(\tau)$ between such pairs. If one of the states in the pair is the ground state $\ket{\psi_0}$, the $\rho_{0j'}(\tau)$ in $\alpha_1(\tau)$ are the time-domain coherences that generate the linear absorption spectrum of the molecule in the frequency domain. However, they are `gated' by the chain of ionization dipole matrix elements $Q_{c0}^{(1)}Q_{cj'}^{(1)*}$ out of the ground state and the wave packet state $\ket{\psi_{j'}}$. In addition to the gated linear absorption spectrum, $\alpha_1(\tau)$ will also contain coherences between any pair of excited states separated by one probe photon in energy.

Similarly, any $\alpha_p(\tau)$ will contain coherences between pairs of wave packet states separated by $p$ probe photons. For $p>1$, if one of the states in the pair is the ground state, then the resulting set of coherences $\rho_{0j'}(\tau)$ generate the $p^{th}$ order Resonance Enhanced Multiphoton Ionization (REMPI) spectrum in the Fourier domain if the resonances are long lived. A set of well chosen $2p_{\text{max}}+1$ measurements of $Y(\tau,\phi)$ thus contain $p_{\text{max}}+1$ different, separable spectroscopic signals. In addition, note from Eq.~\ref{eq:alpha-rho} that each $\alpha_p(\tau)$ can be directly related to a unique set of density matrix elements of the pumped wave packet. This suggests the possibility of determining these matrix elements, or quantum tomography, using the $\alpha_p(\tau)$. 

\section{Illustrative Examples}

In order to illustrate the concepts underlying the technique, we examine two simple cases, each of which serves as a model for the real systems measured in Section~\ref{sec:results}. First we consider an atom or molecule ionized by two photons, with one resonant intermediate state between the ground state $\ket{\psi_0}$ and continuum state $\ket{\psi_c}$ (i.e. $q_{\text{max}}=2$, and the total number of photons absorbed to generate the measured ionization yield is two). Here we do not consider possible two-photon excitation to autoionizing states above the ionization potential. This scenario corresponds to the case of benzene. In this case $-1\leq p \leq 1$, and Eq.~\ref{eq:alpha-signal} may be truncated to three terms. Additionally, since $\alpha_{-p}(\tau)=\alpha_p^*(\tau)$, $Y(\tau,\phi)$ can be expressed in terms of explicitly real quantities:
\begin{equation}
    Y(\tau,\phi)=\Re{\{\alpha_1(\tau)e^{i\phi}\}}+\alpha_0(\tau).
\end{equation}

Each term in this sum results from ionization pathways differing by $1$, or $0$ photons interfering in the continuum. Using the methods discussed in Section~\ref{sec:alphas}, we can determine $\alpha_{1}$ and $\alpha_{0}$ from measurements of $Y(\tau,\phi)$ and Eq.~\ref{eq:alpha-rho}. For the case of benzene, our laser pulses are capable of exciting two vibronic states at the one photon level. In order to describe this as a simple but non-trivial example, we thus take $j=0$ (for the ground state) and $j'$ to be $1a$ or $1b$ (for the two vibronic excited states at the one photon level). The highest term then reduces to:


\begin{equation}
\begin{split}
    \alpha_{1}(\tau) = \sum_{k}W^{(2)}_{c0}W^{(1)*}_{c1a}  Q^{(2)}_{c0}Q^{(1)*}_{c1a}\rho_{01a}(\tau) \\
    +\sum_{k}W^{(2)}_{c0}W^{(1)*}_{c1b}  Q^{(2)}_{c0}Q^{(1)*}_{c1b}\rho_{01b}(\tau)
    \end{split}
\end{equation}
Evidently, the time evolution of $\alpha_1(\tau)$ is determined by coherences between the ground and one-photon resonant excited states ($\rho_{01a}(\tau)$ and $\rho_{01b}(\tau)$). 
The contributions of $\rho_{01a}(\tau)$ and $\rho_{01b}(\tau)$ to the signal are mediated by two transition dipole moments: between the ground state and the one-photon resonant excited states ($1a$ and $1b$), which also determines the absorption spectrum of the system, and from these excited states to the continuum.  Therefore, the excited wave packet responsible for the absorption spectrum is additionally gated by photoionization in this measurement. Again, in the absence of any resonant autoionizing states at the two-photon level, we only expect coherences between one-photon excited states and the ground state to contribute to $\alpha_1(\tau)$.

Finally, we can examine $\alpha_0(\tau)$:
\begin{align}
    \alpha_{0}(\tau) &= \sum_{k=1a,1b}|W^{(2)}_{c0}|^2  |Q^{(2)}_{c0}|^2\rho_{00}(\tau)  \nonumber \\ 
    &+ |W^{(1)}_{c1a}|^2  |Q^{(1)}_{c1a}|^2\rho_{1a1a}(\tau) \nonumber \\
    &+|W^{(1)}_{c1b}|^2  |Q^{(1)}_{c1b}|^2\rho_{1b1b}(\tau) \nonumber \\
    &+W^{(1)}_{c1a}W^{(1)}_{c1b}  |Q^{(1)}_{c1a}||Q^{(1)}_{c1b}|\rho_{1a1b}(\tau).
    \label{eq:zero_photon}
     \end{align}
We note that here $\alpha_0(\tau)$ contains three populations and one coherence between vibronic states excited within the laser bandwidth.  
The information content in $\alpha_0(\tau)$ is similar to that available in TRPES experiments~\cite{stolow2008}, which have been used to track the time-dependent population of individual electronic states. 

A similar analysis can be applied to the case $q_{max}=3$, which allows for $p=0,1$ and $2$. We further simply the discussion here by considering the example of xenon, for which there is no one photon resonance, and we thus only need to worry about $p=0$ and $p=2$. In this case $\alpha_{2}(\tau)$ can each track coherences between two-photon excited states and the ground state, and $\alpha_{0}(\tau)$ will track the populations of the resonantly excited states and coherences between states that are within the bandwidth of the pulse. In the case of the xenon atom, for which data is presented in Section~\ref{sec:results}, two resonant states within the bandwidth of the pulse can be identified from spectral measurements. In this case, we can expect two coherences in $\alpha_2(\tau)$,
\begin{equation}
\begin{split}
\alpha_2(\tau)=\sum_{k}W_{c0}^{(3)}W_{c2a}^{(1)*}Q^{(3)}_{c0}Q^{(1)*}_{c2a}\rho_{02a}(\tau) \\+ \sum_{k}W_{c0}^{(3)}W_{c2b}^{(1)*}Q^{(3)}_{c0}Q^{(1)*}_{c2b}\rho_{02b}(\tau),
\end{split}
\label{eq:alpha2-Xe}
\end{equation}
where $2a$ and $2b$ label the resonantly excited states at the two-photon level. The coherence between the states $2a$ and $2b$ could potentially be measurable in $\alpha_{0}(\tau)$ if both states ionize to the same continuum state. However, in Eq.~\ref{eq:alpha2-Xe}, the continuum state $\ket{\psi_c}$ reached from these two states need \emph{not} be the same. The $\alpha_2(\tau)$ signal instead references each excited state to the ground state, providing an alternative to tracking coherent excited state dynamics when the excited states of interest do not ionize to the same continuum state. This is also true for systems that have multiple resonant states within the bandwidth at the one-photon level. This feature is explicitly demonstrated in Section~\ref{sec:results}.

\section{\label{sec:level1}Experimental Methods}
We performed the measurements using a ytterbium-based laser system (Light Conversion Pharos) producing 255~fs full-width at half-maximum (FWHM) pulses centered at 1025~nm, operating at a repetition rate of $2$-kHz.  The output was converted to its fourth harmonic at 256~nm in the deep ultraviolet (DUV) via two successive second-harmonic generation stages in beta barium borate (BBO) crystals. The pulses were then spectrally broadened through self-phase modulation in a 2~m long krypton-filled stretched hollow-core fiber (200~µm inner diameter; see Ref. \cite{codere2025spectral}) before being compressed and shaped with an acousto-optic modulator (AOM) pulse shaper \cite{dugan1997high, weiner2011ultrafast}. 

The shaper was configured as a 4f zero-dispersion stretcher, with the AOM positioned at the Fourier plane. An arbitrary waveform generator (AWG) drove the AOM transducer to generate an acoustic waveform, which acted as a programmable spectral mask. The initial electric field E($\omega$) is shaped in the frequency domain  by a spectral mask M($\omega$) programmed onto the AOM to create a
pump-probe pulse pair. The shaped electric field E$^\prime$($\omega$) is given by E$^\prime$($\omega$) = M($\omega$)E($\omega$), which allows one to fully control the pump-probe delay $\tau$ and relative phase $\phi$ between the pulses with a mask of the form
\begin{align}
M(\omega)=[1+\exp \left(i\omega \tau + i\phi \right)] \exp(i\varphi(\omega))
\end{align}
where $\varphi(\omega)$ is a phase mask approximated by a Taylor-series expansion up to fourth-order dispersion to minimize the pulse duration in the absence of any other shaping. The compressed $18$-fs DUV pulse pairs are focused into a time-of-flight mass spectrometer (TOF-MS) operating under vacuum. We note that although the pulses themselves are not attosecond-duration pulses, our pulse-shaping technique provides sub-10-attosecond precision in controlling the relative timing between pulses \cite{ngo2026molecular}, comparable to actively stabilized interferometric attosecond beamlines \cite{Koll:22}.

Liquid and gas-phase samples are transferred to a manifold connected to the vacuum chamber. In the case of liquid samples, the manifold is degassed by repeated ``freeze-pump-thaw" cycles to remove air before the sample vapor is introduced into the molecular nozzle. The total ionization yield of benzene (C$_6$H$_6$), acetylene (C$_2$H$_2$), iodomethane-d$_3$ (CD$_3$I) and xenon (Xe) are subsequently measured as a function of the pump-probe delay for several relative phases between the two pulses. Based on the TOF mass spectrum, we have a mass resolution $\frac{m}{\delta m}> 130$. The pulse shaper allows pump-probe measurements to be performed with randomized pump-probe delays.  We make use of delays ranging from 0 to 400~fs in 100~as increments with randomized relative phases between the two pulses. Randomization of the acquisition sequence reduces the impact of slow experimental drifts during data collection and allows for averaging over multiple scans. For each pump-probe delay, we averaged over 200 laser shots.

\section{Results and Discussion}
\label{sec:results}

\begin{figure*}
\includegraphics[width=1\linewidth]{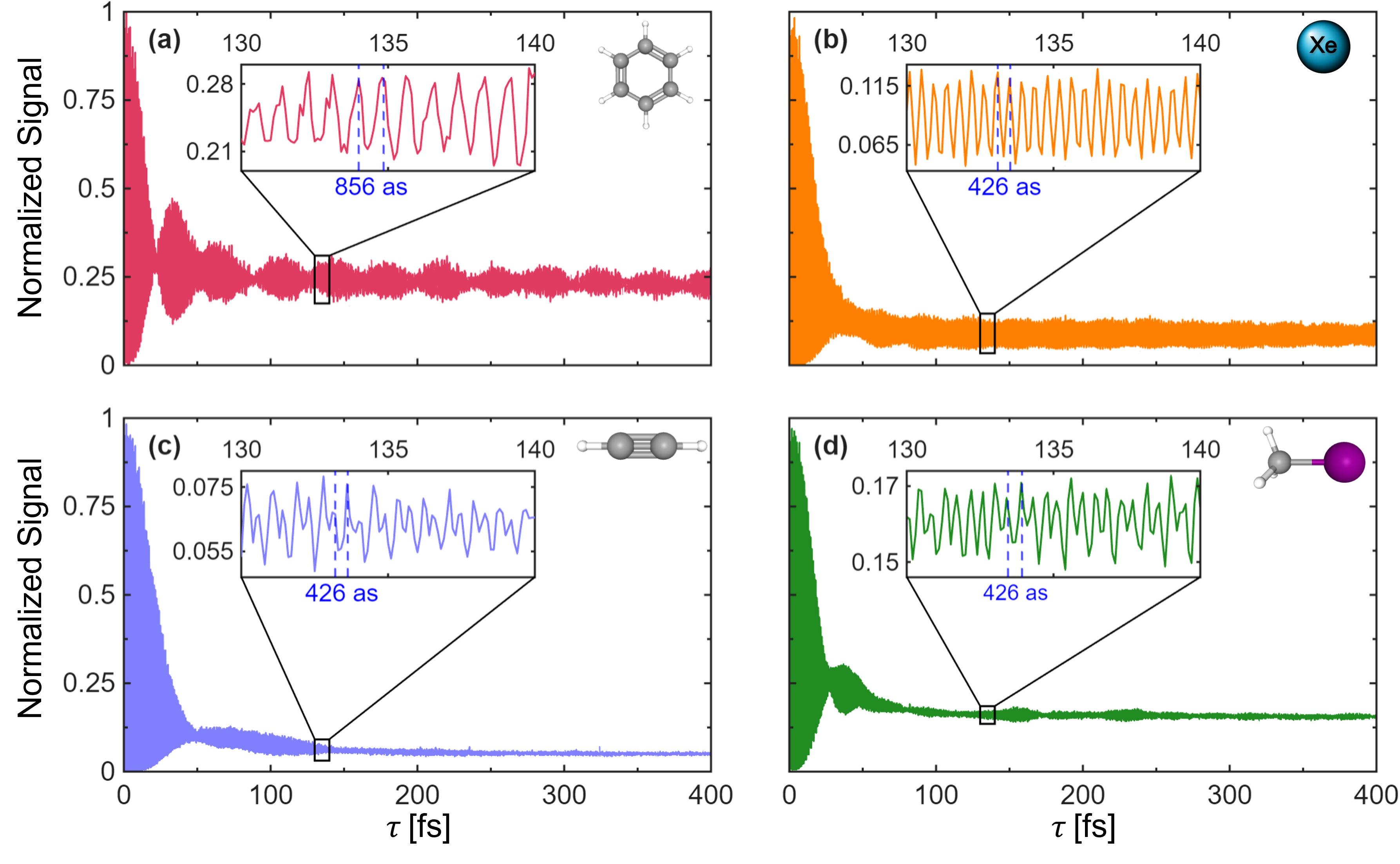} 
\caption{\textbf{Pump--probe measurements for four different systems.}  Each panel shows the parent ion yield as a function of pump-probe delay with a step size of 100 as.  Measurements for four different systems are shown: (Panel a): C$_{6}$H$_{6}$, (Panel b): Xe, (Panel c): C$_{2}$H$_{2}$, (Panel d): CD$_{3}$I. Insets: Zoom of the data over the $130$--$140$~fs range revealing fast oscillations with a period of $856$~as for benzene and $426$~as for xenon. In contrast, both periodicities are observed in acetylene and iodomethane-d$_3$. These periods correspond to one- and two-photon coherences, respectively.}
\label{fig:Time_domain}
\end{figure*}

Figure~\ref{fig:Time_domain} presents the parent ion yields as a function of pump-probe delay for zero phase difference between pump and probe pulses ($\phi=0$) for benzene, xenon, acetylene, and iodomethane-d$_3$ in panels (a)--(d), respectively. The raw signals simultaneously encode information on the populations of individual electronic states and coherences of different photon orders, as written in Eq.~\ref{eq:ptsignalmain}. The insets in each panel show the rapidly oscillating signals as a function of delay $\tau$ that are associated with the electronic beatings for each system. The signal measured in benzene exhibits a quantum beat with a period of 856~attoseconds, while that in xenon displays a distinct beat with a period of 428~attoseconds. In contrast, acetylene and iodomethane-d$_3$ exhibit modulations at both frequencies. This corresponds to  one--photon coherences in benzene, two--photon coherences in xenon and a combination of both in acetylene and iodomethane-d$_3$.

Focusing on panel (a), one can see that the rapid oscillations in benzene are modulated by an envelope with a $36$~fs period.  This is due to the nuclear motion along the symmetric breathing mode of the molecule $\nu_2$ (Herzberg assignments) \cite{ngo2026molecular}. The rapid modulations extend over the entire duration of our pump-probe measurements, highlighting the long-lived coherence between the ground and excited states.  It illustrates how  entanglement between electronic and nuclear degrees of freedom modulates the electronic coherences.  Additionally, these coherences can survive for long times and revive if the dimensionality of the nuclear motion is limited and internal conversion is suppressed (e.g., for excitation below the barrier at wavelengths of about $255$~nm \cite{parker2009ultrafast}).

The xenon measurements shown in panel (b) provide an example of two-photon excitation, leading to long-lived two-photon coherences between a closely spaced pair of excited states and the ground state, while the oscillations in acetylene and iodomethane-d$_3$ in panels (c) and (d) die out earlier due to decoherence.  We note that there is some decay of the signal in xenon, which is due to the finite RF bandwidth in the AOM pulse shaper resulting in a somewhat reduced probe-pulse intensity at large delays.  Interpreting the underlying dynamics from these graphs is challenging because multiple photon-order coherences associated with different excitation pathways are simultaneously present in the ionization yield. 

We used the pulse shaper to perform pump--probe measurements with controlled relative phases between the pulses ($0$, $\pi/4$, $\pi/2$, $3\pi/4$, $\pi$, $-3\pi/4$, $-\pi/2$, and $-\pi/4$). By exploiting the phase dependence of the different photon-order coherences, we can selectively isolate the individual coherence contributions \cite{shim2009turn,myers2008two}. The mathematics describing this phase-cycling  approach and the procedure used to extract the individual photon-order coherences are captured by Eq.~\ref{eq:get_alpha}.\footnote{Another general approach is to compute the Fourier Transform (FT) of the raw signal obtained with a given phase difference (e.g. zero, $\phi=0$) between pulses, apply window functions to the spectral regions corresponding to the 0-, 1-, and 2-photon coherences, and then perform an inverse FT to recover the corresponding coherence contributions as a function of pump--probe delay.}
\begin{figure}[!htb]
\includegraphics[width=1\linewidth]{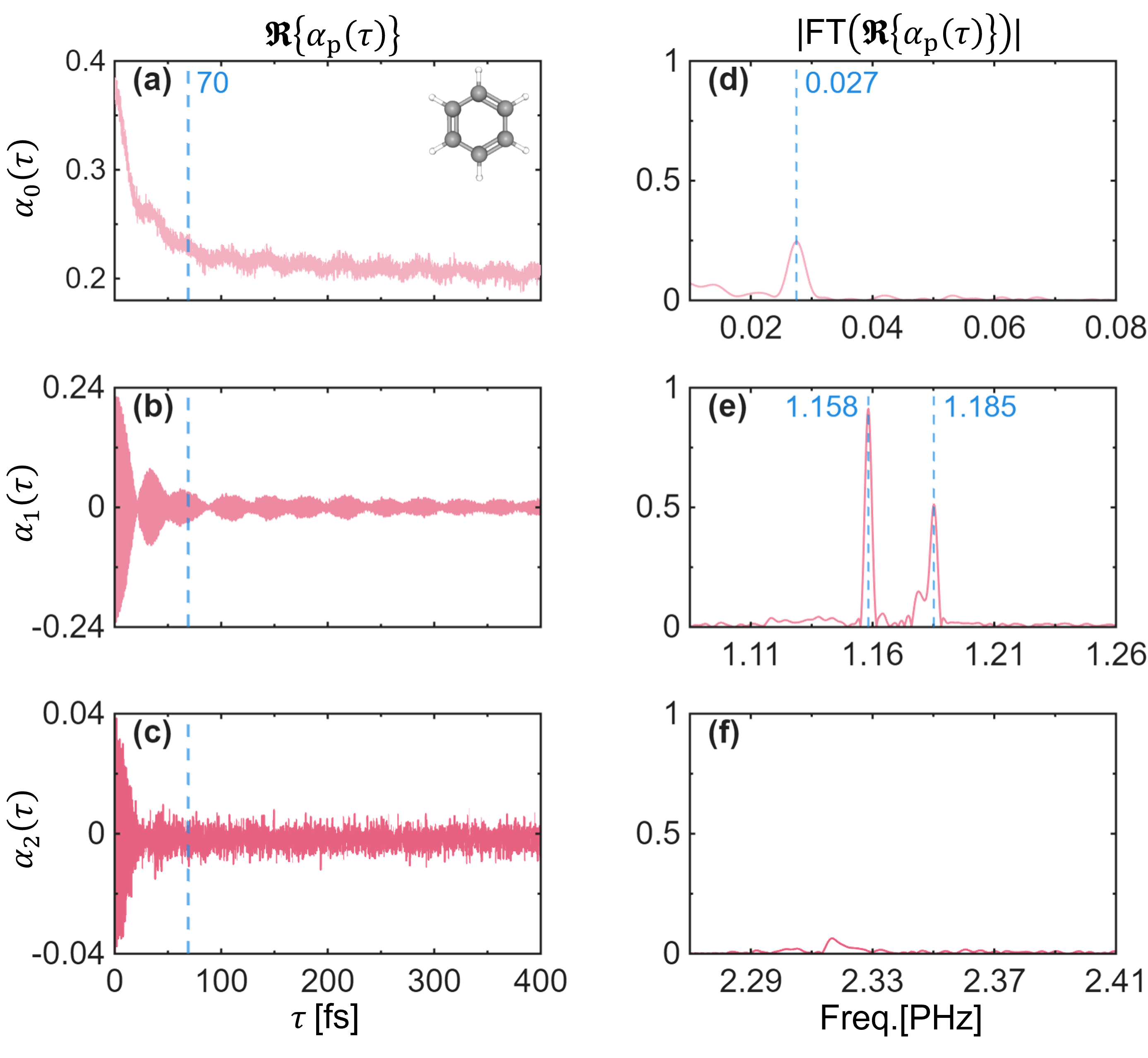}
\caption{\textbf{Time-domain coherence signals  for benzene.} The left column displays the real part of $\alpha_0(\tau)$, $\alpha_1(\tau)$, and $\alpha_2(\tau)$, corresponding to zero-, one-, and two-photon coherences, respectively, as a function of the pump--probe delay (panels~(a)--(c)). The right column shows the Fourier-transform (FT) spectra obtained from the corresponding tim-domain signals truncated below $70$~fs (panels~(d)--(f)). All spectra were normalized to the highest peak in panel (e).}
    \label{fig:BZ_alphas}
\end{figure}

We start with Figure 2, which shows the experimental results of performing the phase-cycling analysis on measurements of benzene. The left column shows the real part of individual coherence contributions as a function of pump-probe delay : (a) $\Re\{\alpha_0(\tau)\}$, (b) $\Re\{\alpha_1(\tau)\}$, and (c) $\Re\{\alpha_2(\tau)\}$. The corresponding FT of the real parts of the individual $\alpha_p(\tau)$ signals are presented in the right column (panels (d)--(f)). We only included the last $\sim$330~fs of the measurements when calculating the FT in order to emphasize the molecular coherence contributions and  avoid contamination by non-resonant optical interference effects when the pump and probe pulses are partially overlapped.

Panel (a) depicts the population of the first neutral excited state together with the associated zero-photon coherence, as described by Eq.~\ref{eq:zero_photon}. Panel (b) displays the one-photon coherence arising from interference between the ground and excited electronic state ionization. The $\alpha_1$ amplitude is modulated by nuclear motion as the excited-state wave packet evolves away from the Franck--Condon region. The underlying dynamics originate from the population of two vibrational eigenstates within the $\nu_2$ progression of the excited electronic $^1B_{2u}$ state following one-photon excitation, resulting in the two peaks observed in the spectrum of panel (e).

Furthermore, the beating between these vibrational eigenstates produces a zero-photon coherence at $\sim 0.027\,\mathrm{PHz}$, also observed in panel (d). It is observable in the delay dependent ionization yield because there can be interference in the final state between the two ionization channels coming from the two vibrational states in the $^1B_{2u}$ electronic state. This interference is possible because both of the vibrational states have the same electronic character and couple to the same continuum state, and because the bandwidth of the probe allows for the contribution of each vibrational state to each of the (limited) vibrational states of the cation that are accessible.   
In contrast, $\alpha_2(\tau)$ [panel (c)] and its FT [panel (f)], show negligible two-photon coherence in benzene.
\begin{figure*}[!htb]
\includegraphics[width=1\linewidth]{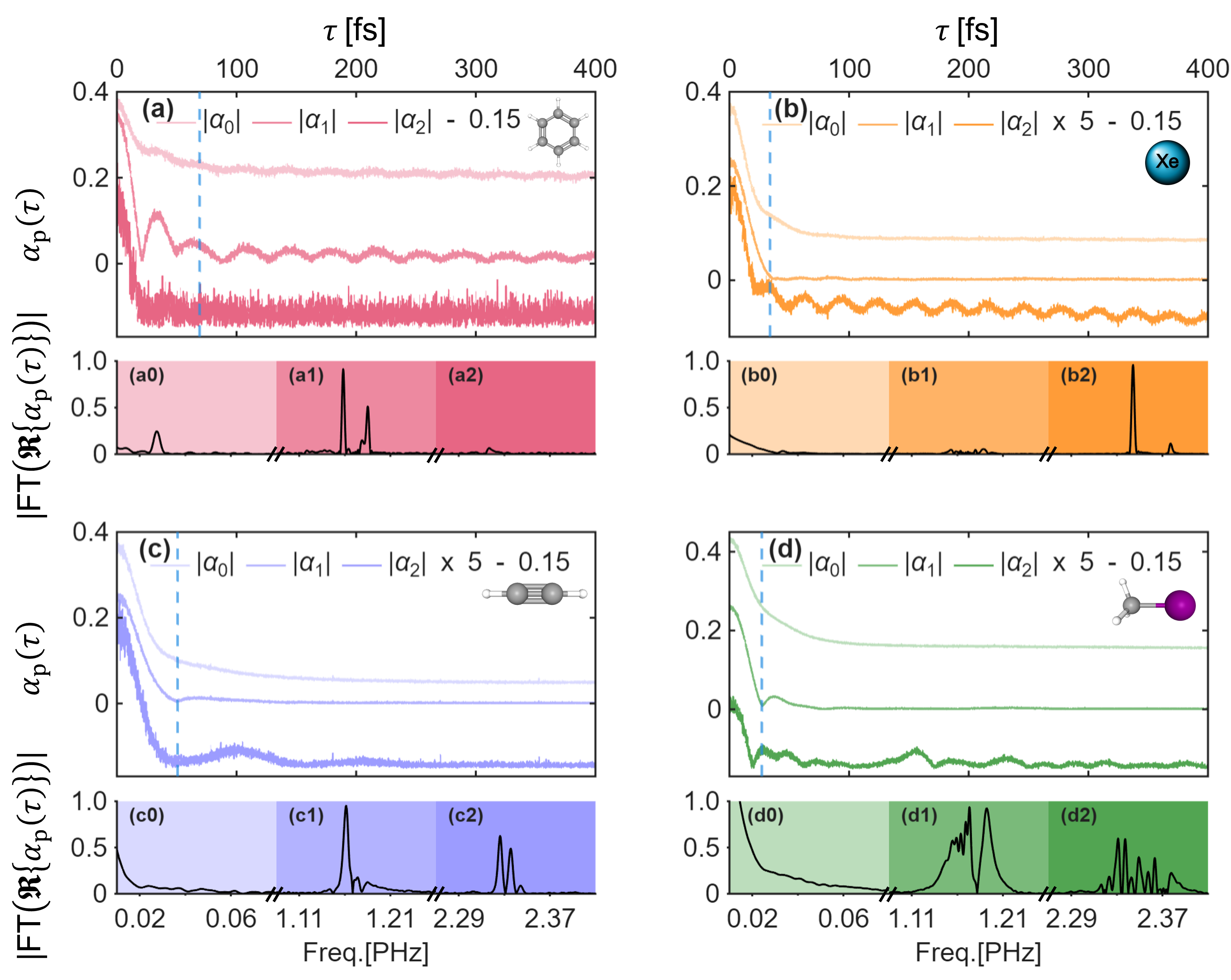}
\caption{\textbf{Time-domain coherence signals and their corresponding FT spectra for four different systems.} The main panels display the magnitudes of the zero-, one-, and two-photon coherences as a function of pump-probe delay for benzene (a), xenon (b), acetylene (c), and iodomethane-d$_3$ (d). Different transparency levels are used to distinguish the coherence orders. The three subpanels beneath each main panel, labeled 0, 1, 2 show the FT spectra of the real parts of $\alpha_0(\tau)$, $\alpha_1(\tau)$ and $\alpha_2(\tau)$  respectively, calculated from the last $\sim$330~fs of the time-domain data (i.e. excluding the pump-probe overlap region near zero time delay, as indicated by the vertical dashed lines). For each system, all FT spectra were normalized to the highest peak among the corresponding subpanels.}
    \label{fig:4systems}
\end{figure*}

The combination of a short pulse duration ($18$~fs) and high enough photon energy ($4.9$~eV per photon) allows us to investigate systems with increasingly complex excited-state dynamics, where both single- and multi-excitation processes may occur. Using an analogous phase-cycling scheme, we extract coherences of different orders, $\alpha_p(\tau)$, from the raw time-domain signals of each system. Panels (a)--(d) of Fig.~{\ref{fig:4systems}} display the magnitudes (in contrast with the real part shown in Fig.~\ref{fig:BZ_alphas}) of the extracted coherence signals for benzene, xenon, acetylene, and iodomethane-d$_3$, respectively. The temporal evolution of these signals provides insight into which states are involved, the coherence lifetimes, revival dynamics, and relative strengths of the various coherence orders supported by each system. The three subpanels beneath each main panel present the Fourier-transform (FT) spectra of the corresponding real parts of the coherence signals, revealing the quantum-beat frequencies associated with the different coherence orders.

\begin{figure}[h]
\includegraphics[width=1\linewidth]{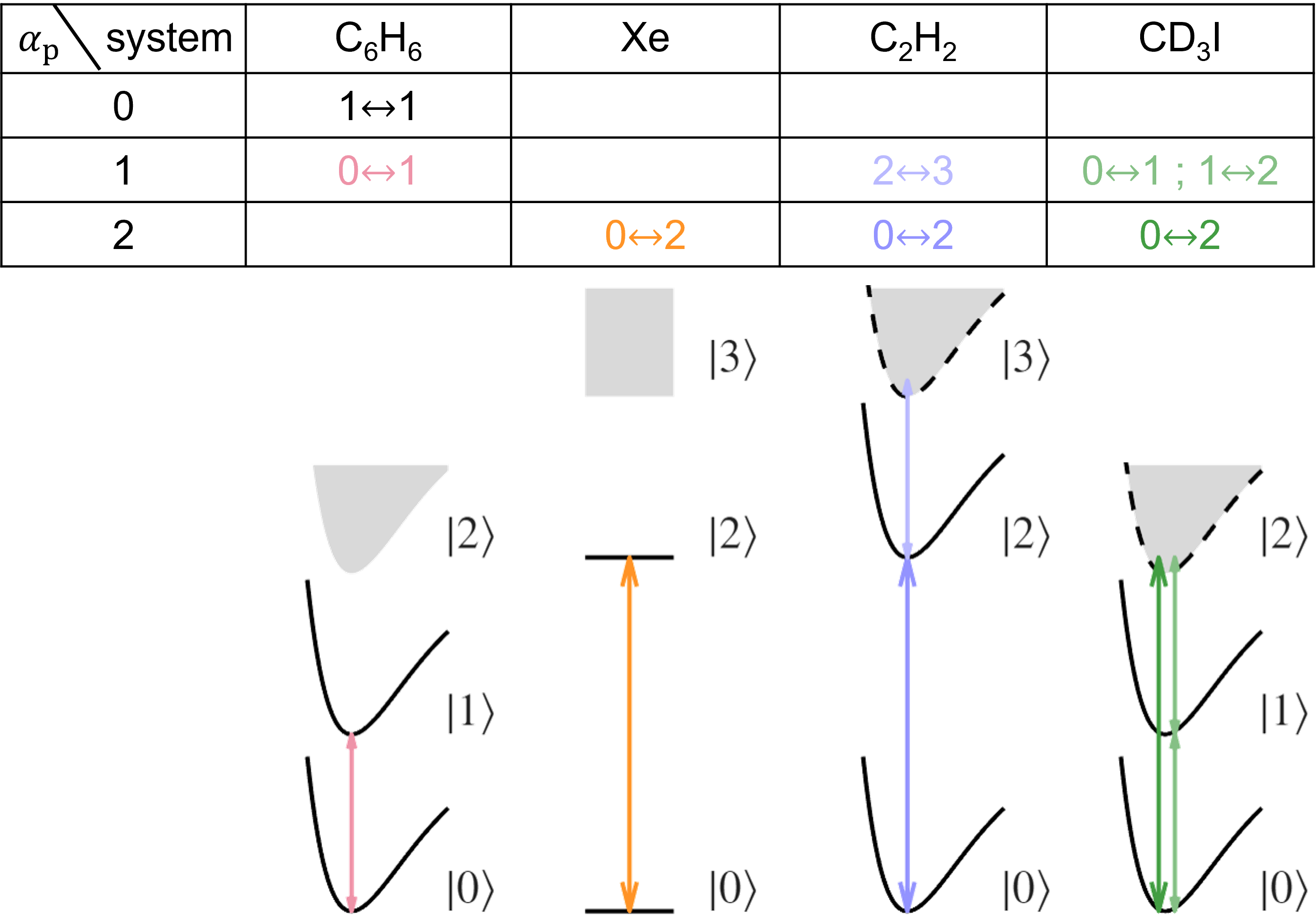}
\caption{\textbf{Electronic coherences of different orders for four systems.} The table summarizes the different coherence orders extracted from the measurements of benzene, xenon, acetylene and iodomethane-d$_3$.  For each system, the table provides the relevant states that are connected by processes of order ``$p$" (via $\alpha_p$). The schematic diagram illustrates the potential energy curves associated with the ground state $\ket{0}$ and the electronically excited states $\ket{1}$, $\ket{2}$, and $\ket{3}$, accessed via one-, two-, and three-photon absorption processes, respectively. The shaded regions denote the ionization continuum corresponding to each system, while the dashed black lines in acetylene and iodomethane-d$_3$ represent autoionizing states lying above the first ionization threshold. Benzene displays one--photon coherence between $\ket{0}$ and $\ket{1}$, and a zero--photon coherence between two vibrational states in state $\ket{1}$. In contrast, xenon, acetylene, and iodomethane-d$_3$ show two-photon coherence between $\ket{0}$ and $\ket{2}$. More intriguingly, acetylene exhibits a one-photon coherence between state $\ket{2}$ and the autoionizing state $\ket{3}$, whereas iodomethane-d$_3$ supports one-photon coherences between states $\ket{0}$ and $\ket{1}$ as well as state $\ket{1}$ and autoionizing state $\ket{2}$.}
    \label{fig:diagram}
\end{figure}

The FT spectra provide a complementary view of the different coherence orders. For benzene (panel (a)), spectral peaks are observed in $\alpha_0$ and $\alpha_1$, while no discernible feature is present in $\alpha_2$, as seen in Fig.~2. In contrast, xenon in panel (b) exhibits two peaks exclusively in $\alpha_2$ (the lower frequency peak is significantly stronger than the high frequency one), with no detectable contributions from $\alpha_0$ and $\alpha_1$. 
Acetylene and iodomethane-d$_3$ (panels (c) and (d) respectively) display spectral features in both $\alpha_1$ and $\alpha_2$, whereas $\alpha_0$ remains featureless except for a peak at zero frequency. These observed coherence-order distributions and their associated state couplings are summarized in Fig.~\ref{fig:diagram}.


Having discussed the benzene measurements above, we now consider electronic coherence in atomic xenon, which serves as a prototype system to extend the measurements beyond one--photon coherences. As shown in panel (b) of Fig.~\ref{fig:4systems}, $\alpha_2(\tau)$ exhibits a pronounced oscillatory behavior that persists over the entire delay range. In panel (b3), the strong peak ($\sim 2.344\,\mathrm{PHz}$) in the FT of $\alpha_2(\tau)$ can be assigned to the two-photon allowed $J=2$ state (the ground state of xenon has $J=0$) with a $5$p$^5$($^2$P$_{3/2}$)$6$p electron configuration with $K=5/2$ (in Racah notation) at $\sim 2.342\,\mathrm{PHz}$~\cite{NIST_ASD}. A much weaker peak in the FT is also observed near $\sim 2.377\,\mathrm{PHz}$ which can be assigned to the $5$p$^5$($^2$P$_{3/2}$)$6$p, $J=2$ state with $K=3/2$ at $\sim 2.375\,\mathrm{PHz}$~\cite{NIST_ASD}. The presence of this additional peak results in the very slight amplitude modulation of the electronic coherence in the time-domain data shown in panel (b) of Fig.~{\ref{fig:Time_domain}}.  This modulation is enhanced and clearly visible in Fig.~\ref{fig:4systems} after extracting the $\alpha_2(\tau)$ contribution.

The electronic dynamics underlying this coherence are easily understood. The superposition of the $J=0$ ground state and $J=2$ excited states generates a time-varying, lab-frame alignment of the electronic probability distribution, where $P_{\text{lab}}(\theta_e,t)=\langle T^2_0\rangle(t)Y_{20}(\theta_e)$, $\theta_e$ is the lab-frame polar angle of the excited electron, and $Y_{20}(\theta_e)$ is the $\ket{l,m}=\ket{2,0}$ spherical harmonic. The quantity $\langle T^2_0\rangle(t)$ is the $\ket{l,m}=\ket{2,0}$ state multipole, as defined by Fano~\cite{fano1957,blum2012density,gregory2022laboratory}, which is proportional to $\langle \cos^2\theta_e\rangle(t)=\mathrm{Tr}\{\hat\rho\cos\theta_e\}$ and follows the time dependence of the coherence between the $J=0$ and $J=2$ states. In contrast to benzene, the FT of $\alpha_0(\tau)$, shown in panel (b1), exhibits no distinct peak. 
Given the fact that both benzene and xenon show $p>0$ coherences with beating in $\alpha_{1,2}$ due to a pair of excited states, it is interesting that benzene also shows a zero-order coherence, while xenon does not.   

In benzene, the two excited states whose zero-photon coherence can contribute to interference in the ionization signal only differ by a vibrational quantum number (they correspond to the same electronic state). This means they can ionize to the same molecular cation and electronic continuum states, producing modulations in $\alpha_0(\tau)$. In contrast, for xenon the two states ionized by the probe pulse correspond to two different electronic states: $5p^5(^2P^0_{3/2})6p$ with $K=3/2$ and $K=5/2$. These states ionize to different electronic continua - continuum states with different energies and angular dependencies.  We estimate that their overlap is less than below 10\%, which explains why there is no observable modulation in $\alpha_0(\tau)$, as seen in panels (b) and (b0) of Fig \ref{fig:4systems} . We argue that this is an important consideration for all ionization-based measurements of electronic dynamics in molecules that aim to probe effective zero-photon coherences. This continuum overlap issue is the continuum analog of Dyson or Koopman's correlations considerations for final molecular cationic states in TRPES measurements \cite{stolow2008}. 

We next consider acetylene, which, similar to xenon, does not possess a resonant one-photon excited state. 
As shown in panel (c3) of Fig. \ref{fig:4systems}, the FT spectrum of $\alpha_2(\tau)$ exhibits three distinct frequency components. Each of these can be assigned to a combination of C$\equiv$C stretch ($\nu_2$) and \textit{cis} bend ($\nu_5$) vibrational states (for brevity, we label these states as $\ket{\nu_2,\nu_5}$, where each digit labels the number of quanta in each mode) in one of two Rydberg electronic states, as shown in Table~\ref{table:Acetylene}. 
The lowest peak corresponds to excitation out of the $\ket{0,1}$ state in the electronic ground state to $\ket{2,2}$ in the $\tilde{D}$ electronic state, while the two higher energy states are both excited out of the $\ket{0,0}$ state in the electronic ground state. $\ket{2,2}$ appears as a weak peak at low temperature, consistent with its assignment to a hot band, whereas $\ket{2,1}$ and $\ket{2,0}$ form a coherent superposition. Since both states in the wave packet possess the same number of stretching quanta, no stretching dynamics are expected. Instead, we probe coherences between the $\ket{0, 0}$ vibrational state in the ground electronic state and the vibronic states $\ket{2, 1}$ $\tilde{D}$, and $\ket{2, 0}$ $\tilde{F}$ as shown in $\alpha_2(\tau)$ in panel (c). Thus this corresponds to an excited-state wave packet consisting of states differing in both electronic and vibrational character. 

\begin{table}[h]
\centering
\caption{Peak assignments for the FT spectrum of acetylene at the two--photon excitation level, based on \cite{jungen1973new}.}
\label{table:Acetylene}
\begin{ruledtabular}
\begin{tabular}{ @{}ccccc@{}}

Frequency & \ $\nu_2$ & \ $\nu_5$ & \ State character & \ State Assignment\\
\midrule
2.326 & 0 $\rightarrow$ 2      & 1 $\rightarrow$ 2 & $3d\sigma, {}^{1}\Pi_u$ & \ $\tilde{D}$ \\
2.336 &  0 $\rightarrow$ 2 & 0 $\rightarrow$ 1 & $3d\sigma, {}^{1}\Pi_u$ & \ $\tilde{D}$  \\
2.344 & 0 $\rightarrow$ 2 &  0 $\rightarrow$ 0 & $nd\pi, {}^{1}\Sigma_u^{+}$ &\ $\tilde{F}$ \\

\end{tabular}
\end{ruledtabular}
\end{table}

The three-photon excitation energy ($\sim 14.5$~eV) falls within a manifold of autoionizing states in acetylene. Previous studies have identified strong Rydberg--valence mixing between the $(3\sigma_g)^{-1}(3p\pi_u)^1$ Rydberg state and the $(3\sigma_g)^{-1}(3\sigma_u)^1$ intravalence  state at about this energy, enhancing coupling to the ionization continuum and thereby facilitating autoionization in this energy region \cite{hattori1997two, hayaishi1982photoionisation, fronzoni2004valence}. Therefore, we assign the one-photon coherence shown in $\alpha_1(\tau)$ in panel (c) of Fig.~\ref{fig:4systems} for acetylene to coherent coupling between the bound neutral and autoionizing states accessed at the two- and three-photon levels, respectively.

Finally, we conclude our analysis with a discussion of iodomethane-d$_3$, which exhibits electronic coherence in both $\alpha_1(\tau)$ and $\alpha_2(\tau)$, but not in $\alpha_0(\tau)$, as shown in panel (d) of  Fig.~\ref{fig:4systems}. As in xenon and acetylene, the FT spectra of iodomethane-d$_3$ reveals the presence of a two-photon coherence; however, the associated modulation pattern is considerably richer and more structured (panel (d3)). The FT uncovers a large number of peaks, six of which can be assigned to different electronic Rydberg states (see Table~\ref{table:cd3i} below). This indicates that there are at least six different electronic states in this region that lie within the two-photon bandwidth of the pulse. In addition, as these states are above the ionization potential of iodomethane-d$_3$, we are resolving coherent electronic dynamics prior to  autoionization of the molecule.
\begin{table}[h]
\centering
\caption{Comparison of peak locations in the FT of our measurements in iodomethane-d$_3$ with the state assignments based on two--photon autoionization/REMPI measurements. The labeling scheme adopted here for Rydberg
states is nl;$\Omega$ where n is the principal quantum number of the Rydberg orbital, l is the orbital angular momentum quantum number, and $\Omega$ is the sum of the axial component of the electronic angular momentum of the ionic core and that of the Rydberg electron.}
\label{table:cd3i}
{\color{black}
\begin{ruledtabular}
\begin{tabular}{ @{}ccc@{}}

Frequency [PHz] & \ REMPI [PHz]   & \ State Assignment\\
\midrule
2.331 & 2.328\footnotemark[1] &7d;$0^{+}$/5f;$0^{+}$ \footnotemark[1]\\
2.337 & 2.334\footnotemark[1] & 9s;1\footnotemark[1]\\
2.343 & 2.343\footnotemark[2]& N/A \\
2.350 & 2.350\footnotemark[2]& 9p$^{\mathrm{a,b}}$\\
2.363 & unresolved & 10s/8d/6f \footnotemark[1] \\
2.380 & 2.380\footnotemark[1] & 10p$^{\mathrm{a,b}}$\\

\end{tabular}
\end{ruledtabular}
\footnotemark[1]{Ref.~\cite{ridley2013critical}.}

\footnotemark[2]{Ref.~\cite{hu2008use}. }
}
\end{table}


Notably, the one-photon coherence in iodomethane-d$_3$ comprises contributions from both the $\ket{0}\leftrightarrow\ket{1}$ and $\ket{1}\leftrightarrow\ket{2}$ coherences, as summarized schematically in Fig.~\ref{fig:diagram}. A detailed interpretation of the underlying dynamics based on electronic-structure calculations will be the subject of a forthcoming manuscript.
Here, we note that all states accessible through two-photon excitation within the bandwidth of our laser pulse lie above the first ionization threshold and are therefore autoionizing. As a result, iodomethane-d$_3$ can be ionized directly without the need for an additional photon, unlike the case of acetylene. Since energy is conserved during the autoionization process (it is simply redistributed amongst the electrons, allowing one to escape), each excited state maps onto a corresponding final ionized state of the same energy. As the excited states are well separated in energy, the resulting final states are likewise distinct and exhibit no overlap. Thus, there is no interference, as seen in the fact that there are no low frequency peaks in the spectrum of $\alpha_0$.

\section{Conclusion}

In conclusion, we have performed and analyzed a series of pump-probe measurements involving both one- and two-photon excitation of electronic wave packets in atomic and molecular systems. The use of phase-locked ultrashort UV pulse pairs, together with independent control of the relative phase and delay, enables direct access to the evolution of distinct coherence orders associated with specific pairs of electronic states. Furthermore, we derive analytical expressions for the ionization yield as a function of pump-probe delay, providing a systematic framework for isolating individual photon-order contributions from the measured signal. Our results demonstrate the ability to isolate and characterize electronic coherences arising from the interference between different excitation pathways and highlights several molecular systems in which we observe long--lived coherences between electronic states over a broad range of energy separations.   \\

The data that support the findings of this study are available
from the corresponding author upon reasonable request.

\begin{acknowledgments}
We gratefully acknowledge Luca Argenti for helpful discussions. 

L.T-H.N., J.C., S.L., S.S., M.G.C., and T.W. were funded by the National Science Foundation under award number 2409596. B.J.P. was funded by Dickinson College. B.K. and M.B. were funded by the Linac Coherent Light Source, SLAC National Accelerator Laboratory, which is supported by the U.S. Department of Energy, Office of Science, Office of Basic Energy Sciences, under Contract No. DE-AC02-76SF00515. R.F. was supported by the U.S. Department of Energy (DOE), Office of Science, Office of Basic Energy Sciences, under Award No. DE-SC0026078.
\end{acknowledgments}
There are no competing interests to declare.

\appendix*

\section{\label{sec:app}Analytic Formulation of the Measurement Observable}

We calculate the ionization amplitude - i.e. the amplitude of the cationic (ionized) state $\ket{\psi_c}$ wave function after the pump and probe pulses. Earlier work made use of a Born-Oppenheimer expansion to relate the measured ionization yield to the underlying wave packet dynamics \cite{ngo2026molecular,kaufman2023long,kaufman2024long}, but here we work in an eigenstate representation. We begin with a Dyson Series expansion of the time propagation operator, $\hat{U}(t) = 1 + \hat{U}^{(1)}(t)+\hat{U}^{(2)}(t)+...$, where
\begin{equation}
\begin{split}
    \hat{U}^{(q)}(t)=&\left(\frac{-i}{\hbar}\right)^q\times\int_{-\infty}^{t}\int_{-\infty}^{t_1}...\int_{-\infty}^{t_{q-1}}dt_1dt_2...dt_{q} \\
    &e^{-i\hat{H}t_1/\hbar}\hat{V}(t_1)e^{i\hat{H}t_1/\hbar} \\
    \times &e^{-i\hat{H}t_2/\hbar}\hat{V}(t_2)e^{i\hat{H}t_2/\hbar}...\\ 
    \times&e^{-i\hat{H}t_q/\hbar}\hat{V}(t_q)e^{i\hat{H}t_q/\hbar}
    \label{eq:dyson}   
    \end{split}
\end{equation}
This operator represents all $q$-photon order interactions of the pumped wave packet with the probe pulse. We insert the dipole potential operator $\hat{V}(t) = \hat{\mu}E(t)$, as well as the unit operator $\sum_{k}\ket{\psi_k}\bra{\psi_k}$ between all propagator pairs.

For a probe pulse that arrives at time $\tau$, we apply $\hat{U}^{q}(t)$ to the wave packet in Eq.~\ref{eq:wkpt-es} to get $A^{(q)}(\tau)$: the $q$-photon order amplitude of ionization into the cationic state $\ket{\psi_c}$,
\begin{equation}
\begin{split}
    &A^{(q)}(\tau)= \bra{\psi_c}\hat{U}^{q}\ket{\psi(\tau)}=\\& e^{iq\phi}\sum_{\gamma}...\sum_{\beta}\sum_{k}\sum_{j}w^{(q)}_{cj}(\Delta E_{c\gamma},...,\Delta E_{\beta k}, \Delta E_{k j})\\  &\times\bra{\psi_c}\hat{\mu}\ket{\psi_\gamma}...\bra{\psi_\beta}\hat{\mu}\ket{\psi_k}\bra{\psi_k}\hat{\mu}\ket{\psi_j}a_{j}e^{-iE_j\tau/\hbar}. \label{eq:nphotonamp}
    \end{split}
\end{equation}
where $w^{(q)}_{cj}$ represents a `window function' (defined shortly in Eq.~\ref{eq:window}).  Each term in the nested sum given in Eq.~\ref{eq:nphotonamp} represents a possible $q$-photon pathway from some state $\ket{\psi_j}$ in the pumped wave packet to the final continuum state $\ket{\psi_c}$ via a set of intermediate states $\{\ket{\psi_\gamma}...\ket{\psi_\beta},\ket{\psi_k}\}$. The number of sums required is determined by the photon order; for example, with $q=2$, there would only be a single sum since with two photons, there can only be a single intermediate step between $\ket{\psi_j}$ and $\ket{\psi_c}$.

For our purposes, the phase $e^{iq\phi}$ of the $q$-photon amplitude in Eq.~\ref{eq:nphotonamp} is the most physically relevant, as it can be experimentally controlled. We can interpret this as single phase factor $e^{i\phi}$ picked up by the absorption of each probe photon on the way to the cation.  We discuss this factor futher after arriving at expressions for the measured signal.

In the `window function' $w^{(q)}_{cj}(\Delta E_{c\gamma},...,\Delta E_{\beta k}, \Delta E_{k j})$, we replace the upper limit of the first integral in Eq.~\ref{eq:dyson} with infinity, implying that the measurement is integrated over all times within the probe pulse. The window function then provides the Fourier components of the pulse at resonant frequencies $\Delta E_{c\gamma}/\hbar,...,\Delta E_{\beta k}/\hbar, \Delta E_{k j}/\hbar$ where $\Delta E_{ij} = E_i - E_j$ is the energy gap between eigenstates. In general,
\begin{equation}
\begin{split}
    w^{(q)}_{cj}(&\Delta E_{c\gamma},...,\Delta E_{\beta k}, \Delta E_{k j})= \\
    &\left(\frac{-i}{\hbar}\right)^q \int_{-\infty}^{\infty}\int_{-\infty}^{t_1}...\int_{-\infty}^{t_{q-1}}dt_1dt_2...dt_{q} \\ 
    &e^{-i(\Delta E_{c\gamma}-\hbar\omega_\ell)t_{q-1}/\hbar}E_0(t_{q})...\\
    \times &e^{-i(\Delta E_{\beta k}-\hbar\omega_\ell)t_{2}/\hbar}E_0(t_{2}) \\
    \times&e^{-i(\Delta E_{k j}-\hbar\omega_\ell)t_{1}/\hbar}E_0(t_{1})
    \end{split}
    \label{eq:window}
\end{equation}
This function essentially enforces the resonance condition at each step of a particular $q$-photon pathway initiating from $\ket{\psi_j}$ and terminating in $\ket{\psi_c}$ (if the step is non-resonant, it provides the appropriate weighting factor).  This can be seen in Eq.~\ref{eq:window} by noting that when the probe photon energy is \textit{not} near resonance with the energy gaps between eigenstates, the detunings (e.g. $\Delta E_{c\gamma}-\hbar\omega_\ell$) in the exponents will be large.  In this case the integral tends to average to zero, and there is a much smaller contribution to the final amplitude.

We can now write the experimental yield as the square of the sum of all possible amplitudes up to some maximum photon order $q_{\text{max}}$.  This yield, which we call $Y(\tau,\phi)$, is simply the $q_{\text{max}}$-photon photoionization probability:
\begin{equation}
Y(\tau,\phi)=|A^{(1)}(\tau,\phi)+A^{(2)}(\tau,\phi)+...+A^{(q_\text{max})}(\tau,\phi)|^2
\label{eq:ptampsum}
\end{equation}
where $q_{\text{max}}$ is the number of photons required to ionize the atom or molecule.  Equation \ref{eq:ptampsum} results in two types of terms: (i) \textit{direct} $q$-photon ionization terms of the form $|A^{(q)}(\tau,\phi)|^2$ that are independent of the phase $\phi$, and (ii) \textit{interference} terms between $q'$ and $q$-photon ionization pathways such as $\left\{A^{(q)}(\tau,\phi)A^{(q')*}(\tau,\phi)+c.c\right\}$.  The latter type depends on the \textit{phase difference} $e^{i(q-q')\phi}$ between the $q$ and $q'$-photon ionization pathways. Each of these terms (i.e. different $p=q-q'$ values) contains complementary information about the pumped wave packet and can be isolated by making the appropriate choices for $\phi$ and adding or subtracting different experimental yields.

In order to see this, we write each delay and $\phi$ dependent term explicitly as follows:
\begin{subequations}   
    \begin{align}
    |A^{(q)}(\tau,\phi)|^2 &= \sum_{\gamma\gamma'}...\sum_{jj'}W^{(q)}_{cj}W^{(q)*}_{cj'}Q^{(q)}_{cj}Q^{(q)*}_{cj'}\nonumber \\
    &\times a_{j}a^{*}_{j'}e^{-i(E_{j} - E_{j'})\tau/\hbar}  \\ 
    A^{(q)}(\tau,\phi)&A^{(q')*}(\tau,\phi) = e^{i(q-q')\phi} \nonumber\\ 
    &\times\sum_{\gamma\gamma'}...\sum_{jj'}W^{(q)}_{cj}W^{(q')*}_{cj'} Q^{(q)}_{cj}Q^{(q')*}_{cj'} \nonumber\\  
    & \times a_{j}a^*_{j'}e^{-i(E_{j} - E_{j'})\tau/\hbar}
    \end{align}
    \label{eq:ptsumterms}
\end{subequations}
where, for brevity, we have made the following notational simplifications for the window functions and transition dipole matrix elements:
\begin{align}
    W^{(q)}_{cj} &\equiv w^{(q)}_{cj}(\Delta E_{c\gamma},...,\Delta E_{\beta k},\Delta E_{k j}) \\
    Q^{(q)}_{cj} &\equiv \bra{\psi_c}\hat{\mu}\ket{\psi_\gamma}...\bra{\psi_\beta}\hat{\mu}\ket{\psi_k}\bra{\psi_k}\hat{\mu}\ket{\psi_j}
    \label{eq:Q_def}
\end{align}

Considering the direct term first, we see that the yield will exhibit quantum beats between those states $\ket{\psi_j}$ and $\ket{\psi_{j'}}$ that can be $q$-photon ionized to the same final ionic state $\ket{\psi_c}$. Because the time-dependent exponent vanishes when $j=j'$, these terms simply provide a DC offset to the yield that is proportional to the populations of these states. Note that for this sum, the $j',j$ terms are the complex conjugates of the $j,j'$ terms (for $j\neq j'$), resulting in a cosine oscillation offset by the relative phases of the window functions, dipole matrix elements, and amplitudes $a_j$. 

The interference terms are similar, although the state $\ket{\psi_j}$ is ionized by $q$ photons, while the state $\ket{\psi_{j'}}$ is ionized $q'$ photons, both ending in the same cationic state $\ket{\psi_c}$. This is not possible if $j=j'$: one eigenstate cannot be ionized by a different number of photons to the same final state. Therefore, this term explicitly rejects populations $|a_j|^2$ and does not contribute to the DC offset of the yield. It only detects quantum beats, or \textit{coherences}, between two states in the pumped wave packet that can be ionized by a different number of photons to the same final state.

With this understanding of the contributions to the yield, we can rewrite Eq.~\ref{eq:ptampsum} in terms of the window functions, transition dipole matrix elements, and the density matrix elements:
\begin{equation}
\begin{split}
&Y(\tau,\phi)= 
\\&\sum_{p=-\infty}^{\infty}e^{ip\phi}\sum_{q\geq p} \sum_{\gamma\gamma'}...\sum_{jj'}W^{(q)}_{cj}W^{(q-p)*}_{cj'}Q^{(q)}_{cj}Q^{(q-p)*}_{cj'}\rho_{jj'}(\tau).
\label{eq:ptsignal}
\end{split}
\end{equation}
where $\rho_{jj'}(\tau)=a_ja_{j'}^*e^{-i(E_{j} - E_{j'})\tau/\hbar}$ is the density matrix of the pumped wave packet at a time delay $\tau$ in the eigenstate basis. This corresponds to Eq.~\ref{eq:ptsignalmain} in the main text.
\vspace{20pt}
\nocite{*}

\bibliography{references}

\end{document}